\documentclass[aps,pra,twocolumn,showpacs]{revtex4-1}

\usepackage{amsfonts,amssymb,amsmath}
\usepackage{mathtools}
\usepackage[]{graphics,graphicx,epsfig}
\usepackage{amsthm,multirow}
\usepackage{color}
\begin{document}

\title{Quantum illumination via quantum-enhanced sensing}

\author{Su-Yong \surname{Lee}}
\affiliation{Quantum Physics Technology Directorate, Agency for Defense Development, Daejeon 34186, Korea}

\author{Yong Sup Ihn}
\affiliation{Quantum Physics Technology Directorate, Agency for Defense Development, Daejeon 34186, Korea}

\author{Zaeill Kim}
\affiliation{Quantum Physics Technology Directorate, Agency for Defense Development, Daejeon 34186, Korea}

\date{\today}

\begin{abstract}
Quantum-enhanced sensing has a goal of enhancing a parameter sensitivity with input quantum states, while quantum illumination has a goal of enhancing a target detection capability with input entangled states in a heavy noise environment.
Here we propose a concatenation between quantum-enhanced sensing and quantum illumination that can take quantum advantage over the classical limit.
First, phase sensing in an interferometry is connected to a target sensing via quantum Fisher information. Second, the target sensitivity is investigated in noisy quantum-enhanced sensing. Under the same input state energy, for example, $N$-photon entangled states can exhibit better performance than a two-mode squeezed vacuum state and a separable coherent state.
Incorporating a photon-number difference measurement, finally, the noisy target sensitivity is connected to a signal-to-noise ratio which is associated with a minimum error probability of discriminating the presence and absence of the target. We show that both the target sensitivity and the signal-to-noise ratio can be enhanced with increasing thermal noise.
\end{abstract}

\maketitle

\section{Introduction}
Quantum-enhanced sensing(QES) takes quantum advantage over classical strategies via input entanglement and squeezing\cite{giovannetti2004,dowling2008,Stefano18}. 
For a single-parameter sensing, a parameter sensitivity is lower bounded by the inverse of quantum Fisher information(QFI)\cite{BC94} which provides maximum information we can extract in a small change of the parameter, where the mean value of the parameter is equal to the true value of the parameter.
According to input states, the sensitivity of a phase is lower bounded by the standard quantum limit ($1/\sqrt{N}$) with coherent states or by the Heisenberg limit ($1/N$) with $NOON$ states and squeezed states, where $N$ is a mean photon number of an input state.
In a noisy scenario, we can explore noisy quantum-enhanced sensing\cite{Escher} as well as quantum illumination(QI)\cite{Lloyd,Tan,Shapiro} that discriminates the presence and absence of a target. For the QI which detects the target with entangled states in a heavy noise environment, the target is simply modeled by a beam splitter in a laboratory.
Using entangled states, we can enhance the possibility of detecting the target even if there is no entanglement in the output modes. Specifically, two-mode squeezed vacuum(TMSV) states can exhibit quantum advantage over classical states in QI\cite{Tan} with no output entanglement, where it was not shown how to achieve the quantum advantage with any measurement setup.

QI using an input TMSV state was first implemented with direct photon counting on output modes\cite{Genovese}, where the signal-to-noise ratio(SNR) of the TMSV state was higher than the SNR of a correlated thermal state.
Regarding a specific measurement setup, Guha and Erkmen proposed a measurement scheme with an optical parametric amplifier that achieves a half of the bound\cite{Guha}. It was implemented by Zhang et al.\cite{Zheshen}, showing the $20\%$ improvement of the SNR over optimal classical setup.
Later, Zhuang, Zhang, and Shapiro proposed another measurement scheme\cite{Quntao} in which sum-frequency generation with feedforward can achieve the bound asymptotically. 
In addition to those works, there were several investigations on the QI both theoretically\cite{SL09,Devi,Ragy,Zhang14,Shabir,Sanz,Liu17,Weedbrook,Bradshaw,Zubairy,Stefano19,Palma,Yung,Ray,Sun,Ranjith} and experimentally\cite{Shabir19,Sandbo,England,Aguilar,Sussman}.

\begin{figure}
\centerline{\scalebox{0.65}{\includegraphics[angle=0]{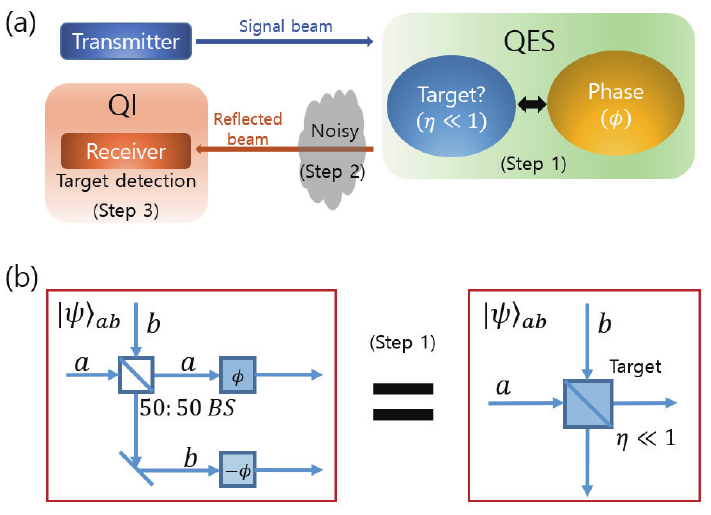}}}
\caption{(a) Schematic for quantum illumination(QI) via quantum-enhanced sensing(QES), which consists of three steps.
(b) (Step 1) Equivalence between sensing a phase($\phi$) and sensing a target reflectivity($\eta\ll 1$) via quantum Fisher information.
(Step 2) Noisy QES in Fig. 2, and (Step 3) Noisy QES is adapted for QI which is based on a target detection with SNR in Fig. 3.}
\label{fig:fig1}
\end{figure}

Both QI and QES take quantum advantage with input entangled states, but the former can achieve it without output entanglement\cite{Tan} and the latter can do it with output entanglement\cite{Lee}. Moreover the QI is based on quantum discrimination and the QES is based on quantum estimation\cite{Stefano18}.
Although they exhibit different characteristics,  the QI and the QES can be connected quantitatively by a target sensitivity.
In Fig. 1 (a), we draw a schematic for concatenating QI and QES. 

\section{Equivalence between different parameter sensitivities}
The most well-known QES is to estimate a phase with an unbiased estimator whose mean value is equal to the true value of the phase, where the phase sensitivity is induced by a path length difference. Including a $50:50$ beam splitter, we show that the phase sensing can be equivalent to a target sensing with a low reflectivity via QFI.
For pure states, the QFI is given by $H=4[(\frac{\partial_{ab}\langle \psi_x|}{\partial x})(\frac{\partial |\psi_x\rangle_{ab}}{\partial x})-|_{ab}\langle\psi_x|\frac{\partial |\psi_x\rangle_{ab}}{\partial x}|^2]$, where $|\psi_x\rangle_{ab}=\hat{U}_{ab}(x)|\psi\rangle_{ab}$.
First, we look into the target which is replaced by a beam splitter. 
A general beam splitting operation\cite{BS,Kim} is represented by
\begin{eqnarray}
\hat{B}_{ab}(\theta,\varphi)&=&\exp\bigg[\frac{\theta}{2}(\hat{a}^{\dag}\hat{b}e^{i\varphi}-\hat{a}\hat{b}^{\dag}e^{-i\varphi})\bigg]\nonumber\\
&\approx & \exp[\eta(\hat{a}^{\dag}\hat{b}e^{i\varphi}-\hat{a}\hat{b}^{\dag}e^{-i\varphi})]\equiv \hat{B}_{ab}(\eta,\varphi),
\end{eqnarray}
where $\eta=\sin(\theta/2)$ is the reflectivity of a beam splitter, and $\varphi$ is the phase difference between the transmitted and reflected fields.
Assuming $\theta \ll 1$, the beam splitting operation is approximated as $\hat{B}_{ab}(\eta,\varphi)$.
Applying the beam splitting operation on a two-mode input state as $|\psi_{\eta}\rangle_{ab}=\hat{B}_{ab}(\eta,\varphi)|\psi\rangle_{ab}$, 
we derive the QFI
\begin{eqnarray}
H&=&-4\bigg[ ~_{ab}\langle\psi |(\hat{a}^{\dag}\hat{b}e^{i\varphi}-\hat{b}^{\dag}\hat{a}e^{-i\varphi})^2|\psi\rangle_{ab}\nonumber\\
&&+\bigg| ~_{ab}\langle\psi |(\hat{a}^{\dag}\hat{b}e^{i\varphi}-\hat{b}^{\dag}\hat{a}e^{-i\varphi})|\psi\rangle_{ab}\bigg|^2\bigg].
\end{eqnarray}
At $\varphi=\pi/2$, the QFI is given by $H=4[\langle (\hat{a}^{\dag}\hat{b}+\hat{b}^{\dag}\hat{a})^2\rangle-|\langle \hat{a}^{\dag}\hat{b}+\hat{b}^{\dag}\hat{a}\rangle |^2]$.
We can derive the same QFI formula for sensing a phase in interferometry, as shown in Fig. 1 (b).  
A two-mode input state impinges on a $50:50$ beam splitter, and then experiences a phase shifter with an opposite sign on each arm.
Using the $50:50$ beam splitting operation of $\hat{a}^{\dag}\rightarrow \frac{1}{\sqrt{2}}(\hat{a}^{\dag}-ie^{-i\varphi}\hat{b}^{\dag})$ and
 $\hat{b}^{\dag}\rightarrow \frac{1}{\sqrt{2}}(\hat{b}^{\dag}-ie^{i\varphi}\hat{a}^{\dag})$,
we derive the output state 
\begin{eqnarray}
|\psi_\phi\rangle_{ab}&=&e^{i\phi(\hat{a}^{\dag}\hat{a}-\hat{b}^{\dag}\hat{b})}\hat{B}_{ab}(\frac{\pi}{2},\varphi+\frac{\pi}{2})|\psi\rangle_{ab}\nonumber\\
&=&\hat{B}_{ab}(\frac{\pi}{2},\varphi+\frac{\pi}{2})e^{i\phi(\hat{a}^{\dag}\hat{b}e^{i\varphi}-\hat{a}\hat{b}^{\dag}e^{-i\varphi})}|\psi\rangle_{ab},
\end{eqnarray}
where we employ the unitary relation of the beam splitting operation $\hat{B}^{\dag}_{ab}\hat{B}_{ab}=\hat{B}_{ab}\hat{B}^{\dag}_{ab}=\hat{I}$.
 Then, the associated QFI becomes the same as Eq. (2).
In an ideal scenario, thus, sensing a phase($\phi$) is equivalent to sensing a target($\eta\ll 1$) via QFI. 
Given a unitary operation $\hat{U}_{ab}(x)=e^{ix\hat{O}_{ab}}$, in general, we can derive the similar relation if the other unitary operation is transformed into $\hat{U}_{ab}(y)=\hat{A}e^{iy\hat{O}_{ab}}$, where $\hat{A}$ is independent of a parameter $y$.
The sensitivity of $x$ is equivalent to the sensitivity of $y$ by their QFI formula.

When one of the input modes is a coherent state such as $|\psi\rangle_{ab}=|\alpha\rangle_a|\Psi\rangle_b$, the QFI of Eq. (2) is given by
\begin{eqnarray}
H_{|\alpha\rangle_a|\Psi\rangle_b}=4\bigg[\langle \hat{b}^{\dag}\hat{b}\rangle+2|\alpha|^2\Delta X^2_{\Theta+\pi/2}\bigg],
\end{eqnarray}
where $\Delta X^2_{\Theta+\pi/2}=\langle \hat{X}^2_{\Theta+\pi/2}\rangle-|\langle \hat{X}_{\Theta+\pi/2}\rangle|^2$ is the variance of a quadrature operator $\hat{X}_{\Theta+\pi/2}=(\hat{b}e^{-i\Theta}-\hat{b}^{\dag}e^{i\Theta})/i\sqrt{2}$. Note that $\alpha=|\alpha|e^{i\theta}$ and $\Theta=\theta-\varphi$. Since the optimal condition of the other input mode is antisqueezed in the direction of $\Theta+\pi/2$,
it is best to inject a squeezed vacuum state in the input mode $b$ \cite{Lang13}.
When one of the input modes is a vacuum state such as $|\psi\rangle_{ab}=|0\rangle_a|\Psi\rangle_b$, the QFI is given by $4\langle \hat{b}^{\dag}\hat{b}\rangle$ such that the optimal condition is proportional to the mean photon number of the other input mode.
The best thing is to inject a coherent state in the input mode $b$.

\section{noisy quantum-enhanced sensing}

\begin{figure}
\centerline{\scalebox{0.7}{\includegraphics[angle=0]{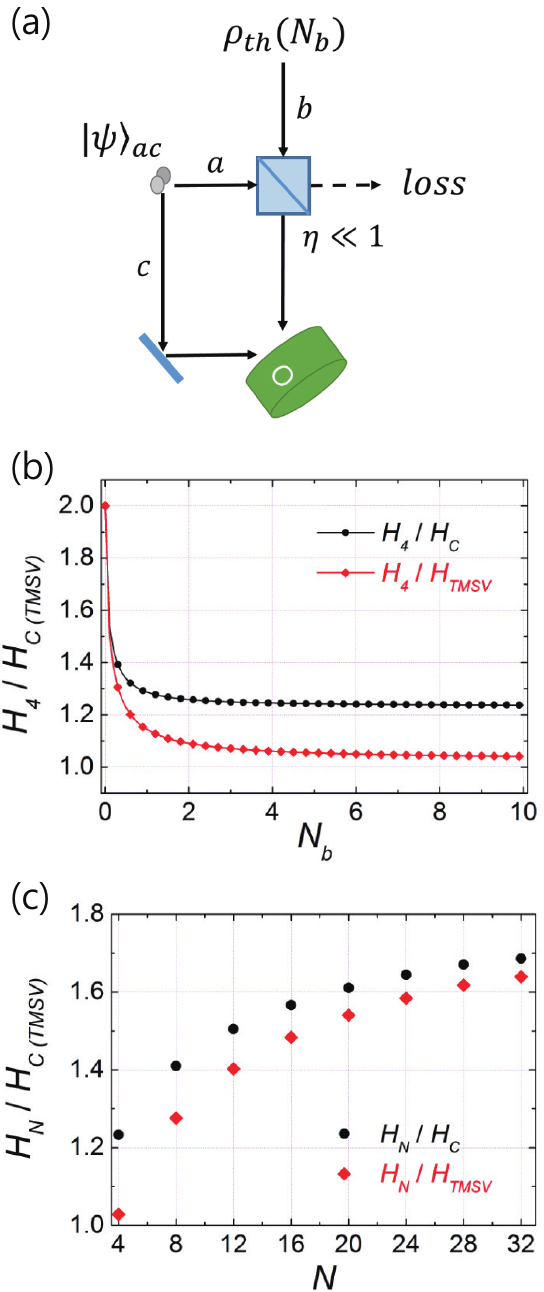}}}
\caption{(a) Noisy quantum-enhanced sensing on a target reflectivity.
$O$ represents an optimal measurement, and $\rho_{th}$ is a thermal noise with mean photon number $N_b$.
Under the constraint of input energy $\langle \hat{n}_a+\hat{n}_c\rangle=N$,
we show (b) the gain in the QFI as a function of $N_b$:
a $4$-photon entangled state versus a TMSV state (red curve with squares), or a separable coherent state (black curve with circles) that is considered as a reference state. (c) Gain in the QFI as a function of $N$ from the $N$-photon entangled state, at $N_b=10$, which is almost the saturating point.
}
\label{fig:fig2}
\end{figure}

Including thermal noise and loss, we manipulate the target sensitivity in noisy quantum-enhanced sensing.
For mixed states, we utilize the QFI formula\cite{Paris} that is given by $H=\sum_{nm}\frac{2|\langle\phi_n|(\partial_{\eta}\rho_{\eta})|_{\eta=0}|\phi_m\rangle|^2}{(\lambda_n+\lambda_m)}$, where $(\partial_{\eta}\rho_{\eta})$ is the derivative of the output state $\rho_{\eta}$. $\lambda_m$ and $|\phi_m\rangle$ are the eigenvalues and the eigenstates of $\rho_{\eta=0}$, respectively.
Note that a higher QFI represents a better target sensitivity.
In Fig. 2 (a), we insert a thermal state into the input mode $b$ as a thermal noise effect, and one of the output modes is discarded.
Based on the QFI of $\eta(\ll 1)$\cite{Sanz}, 
we consider $N$-photon entangled states\cite{Rafal} which are given by the formula, $\sum^N_{n=0}a_n|N-n,n\rangle_{ac}$, where $\sum^N_{n=0}|a_n|^2=1$. 
We describe the generation scheme of the $N$-photon entangled states in Appendix A.
 In the constraint of a total input state energy, we show that $N$-photon entangled states can beat the performance of  a TMSV state regardless of the amount of thermal noise. 
The gain in the QFI of the $N$-photon entangled state versus the QFI of the TMSV state increases with $N$, where
the coefficients of the $N$-photon entangled states are optimized.

The QFI of the $N$-photon entangled state is derived as
\begin{eqnarray}
H_{N}=\frac{4}{1+N_b}\sum^{N-1}_{n=0}\frac{(n+1)a^2_{n+1}a^2_n}{a^2_n+a^2_{n+1}(\frac{N_b}{1+N_b})},
\end{eqnarray}
where $N_b$ is the mean photon number of thermal noise.
At $N_b=0$, all the photons of the $N$-photon entangled state are located on the signal mode as $|N,0\rangle_{ac}$ which is an optimal state for the vacuum noise. 
In a range of $N\geq 4$, we obtain that numerically the QFI of the $N$-photon entangled state is larger than the QFI of the TMSV state, irrespective of $N_b$.  
 The QFIs of a coherent state and a TMSV state are given by $H_C=\frac{2N_s}{2N_b+1}$ and
$H_{\text{TMSV}}=\frac{2N_s}{(N_b+1)[1+(\frac{N_s}{N_s+2})(\frac{N_b}{N_b+1})]}$ respectively, where $N_s=\langle \hat{n}_a+\hat{n}_c\rangle$ is the total input mean photon number.
Since the TMSV state and the separable coherent state($|\alpha\rangle_a|\alpha\rangle_c$) are equally distributed in both signal and idler modes, the QFI of the $N$-photon entangled state is twice of the QFI of the other states at $N_b=0$.
In Fig. 2 (b), we observe that, at $\langle \hat{n}_a+\hat{n}_c\rangle=4$,  the QFI of a four-photon entangled state is about $1.24 (1.04)$ times larger than the QFI of the separable coherent(or TMSV) state at $N_b=10$.
The corresponding coefficients are given in Appendix B.
Correspondingly, the mean photon number of the signal ($\langle \hat{n}_a\rangle$) decreases from $4$ to $3$ with $N_b$ whereas that of the idler ($\langle\hat{n}_c\rangle$) increases from $0$ to $1$.
This implies that, given a same input state energy, asymmetric entangled states can show better performance than symmetric entangled states in the target sensitivity. At $N_b=10$, which is almost the saturation point of the QFIs, the gain in the QFI of an $N$-photon entangled state over the QFI of the separable coherent(or TMSV) state increases with the number $N$, as shown in Fig. 2 (c). 


The target sensitivity is lower bounded by $\Delta\eta\geq1/\sqrt{mH}$, where $H$ is the QFI and $m$ is the photon flux. Thus, actual enhancement is represented with the QFI multiplied by the photon flux. Under the same photon flux, e.g., $m=10000$, for comparison the amount of gain is obtained as $m(H_{\text{N}}-H_{\text{TMSV}})=6000$ at $N=32$ and $400$ at $N=4$. The enhancement increases with the photon flux.

Since the QFI of the $N$-photon entangled state requires the corresponding correlated measurement that is derived with the symmetric logarithmic derivative of $\rho_{\eta}$ calculated at $\eta=0$\cite{Paris}, it is hard to implement in a laboratory. In the next part, we propose a feasible measurement setup to connect target sensitivity and target detection.

\section{Sensing a target with a specific measurement setup}

\begin{figure}
\centerline{\scalebox{0.8}{\includegraphics[angle=0]{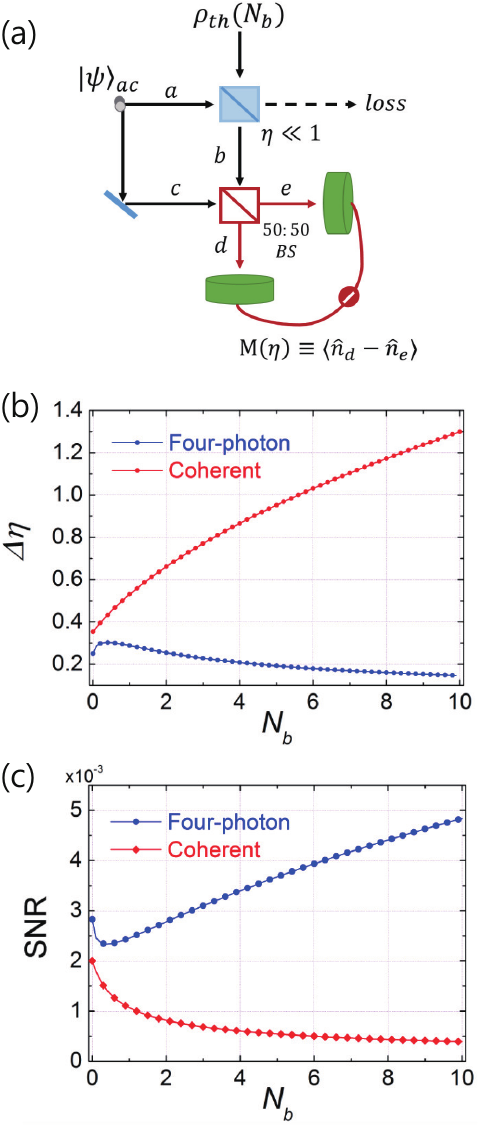}}}
\caption{(a) Measurement setup, (b) target sensitivity, and (c) signal-to-noise ratio for the reflected signal and the idler modes, as a function of the mean photon number of thermal noise:
a four-photon entangled state (blue curve with circles) and a separable coherent state (red curve with squares) at $\eta=10^{-3}$ and $\langle \hat{n}_a+\hat{n}_c\rangle=4.$
The $50:50$ beam splitter results in the transformation of $\hat{d}^{\dag}\rightarrow \frac{1}{\sqrt{2}}(\hat{b}^{\dag}+\hat{c}^{\dag})$ and $\hat{e}^{\dag}\rightarrow \frac{1}{\sqrt{2}}(\hat{c}^{\dag}-\hat{b}^{\dag})$. }
\label{fig:fig3}
\end{figure}

We consider an implementable measurement setup for target sensitivity and target detection in a noisy environment. 
The target sensitivity is simply evaluated with the error propagation relation 
$\Delta\eta=\sqrt{\Delta M(\eta)^2}/|\frac{\partial M(\eta)}{\partial \eta}|$, where
$\Delta M(\eta)^2=\langle\hat{M}^2\rangle-\langle\hat{M}\rangle^2$ and $\langle\hat{M}\rangle=M(\eta)$.
The target detection is defined as $\text{SNR}\equiv[M(\eta)-M(0)]/[\sqrt{\Delta M(\eta)^2}+\sqrt{\Delta M(0)^2}]$.
Combining both formulas, we derive the following relation
\begin{eqnarray}
\text{SNR}=\frac{M(\eta)-M(0)}{\Delta \eta \bigg|\frac{\partial M(\eta)}{\partial \eta}\bigg|+\sqrt{\Delta M(0)^2} }.
\end{eqnarray}
Given a $M(\eta)$, we observe that the SNR is inversely proportional to $\Delta \eta$.
We can infer that the better the target sensitivity is, the more probable the target detection is.

The SNR is associated with a minimum error probability of distinguishing between the presence and absence of a target. In the Gaussian regime, the minimum error probability is given by $e^{-MR_G}/2\sqrt{\pi MR_G}$ \cite{Guha}, where $M(\gg 1)$ is the number of pairs for returned and idler modes, 
and $R_G=(\overline{n}_1-\overline{n}_0)^2/2(\sigma_0+\sigma_1)^2$ is the error exponent. $\overline{n}_1(\overline{n}_0)$ is the mean photon number of the presence(absence) of the target. 
$\sigma_1(\sigma_0)$ is the standard deviation of the presence(absence) of the target. In our non-Gaussian regime, we simply define the error exponent as 
$R_{nG}\equiv (\overline{n}_1-\overline{n}_0)^2/(\sigma_0+\sigma_1)^2 \equiv \text{SNR}^2$, where $\text{SNR}=\sqrt{R_{nG}}$ we call an effective SNR.

We assume that an input thermal noise is separately distributed after a beam splitting operation as $ \rho^{(a)}_{th}(\eta^2N_b)\otimes \rho^{(b)}_{th}((1-\eta^2)N_b)$ in the condition of 
$\eta^2N_b \ll 1$ which satisfies the Gaussian R\'enyi-2 mutual information\cite{Adesso} of the output thermal noise being $I_2=\ln[1+2\eta^2(1-\eta^2)\frac{N^2_b}{1+2N_b}]\approx 0$.


In Fig. 3  (a), we consider a photon-number difference measurement after combining the reflected signal and the idler by a 50:50 beam splitter whose transformation is given by $\hat{d}^{\dag}\rightarrow \frac{1}{\sqrt{2}}(\hat{b}^{\dag}+e^{i\varphi}\hat{c}^{\dag})$ and $\hat{e}^{\dag}\rightarrow \frac{1}{\sqrt{2}}(\hat{c}^{\dag}-e^{-i\varphi}\hat{b}^{\dag})$. Using the photon-number difference measurement, we can distinguish the presence of a target from the absence of one, distinctly. 
In the absence of the target, the output state ($\rho^{(b)}_{th}\otimes tr_a[|\psi\rangle_{ac}\langle \psi|]$) is always observed  as 
$M(\eta)=\langle \hat{b}^{\dag}\hat{c}e^{-i\varphi}+\hat{c}^{\dag}\hat{b}e^{i\varphi} \rangle=0$, 
in which the measurement observable $\hat{M}= \hat{n}_d-\hat{n}_e$ is transformed into 
$(\hat{b}^{\dag}\hat{c}e^{-i\varphi}+\hat{c}^{\dag}\hat{b}e^{i\varphi})$ by the reverse 50:50 beam splitting operation.
In the presence of the target, we measure the interference terms in the modes $b$ and $c$.
However, there is an exception that $M(\eta)=0$ can indicate the presence of the target with an input pure two-mode Gaussian state having zero first moments, since the off-diagonal elements of the output covariance matrix have the relation \cite{Adesso04}, $\langle \hat{X}_b\hat{X}_c\rangle=-\langle \hat{P}_b\hat{P}_c\rangle$ and $\langle \hat{X}_b\hat{P}_c\rangle=\langle \hat{P}_b\hat{X}_c\rangle=0$.
The measurement observable is reformulated with the corresponding position and momentum operators, $\langle \hat{b}^{\dag}\hat{c}e^{-i\varphi}+\hat{c}^{\dag}\hat{b}e^{i\varphi} \rangle=\langle (\hat{X}_b\hat{X}_c+ \hat{P}_b\hat{P}_c)\cos{\varphi}+(\hat{X}_b\hat{P}_c- \hat{P}_b\hat{X}_c)\sin{\varphi}\rangle$.
For that reason, we do not consider a TMSV state under the photon-number difference measurement.

Based on the photon-number difference measurement, we explore whether entangled states can exhibit better performance than a separable coherent state.
For example, we consider a four-photon entangled state in comparison with the separable coherent state, at the low reflectivity of $\eta=10^{-3}$.
For the target sensitivity, the four-photon entangled state initially deteriorates with $N_b$ but improves 
from $N_b\approx 0.5$, as shown in Fig. 3 (b). 
Meanwhile the performance of the separable coherent state deteriorates with $N_b$.
For the SNR, correspondingly, the four-photon entangled state exhibits a similar tendency, as shown in Fig. 3 (c). 
With increasing thermal noise, at $\langle \hat{n}_a+\hat{n}_c\rangle=4$,
 the mean photon number of the signal($\langle \hat{n}_a\rangle$) decreases from $4$ to $3$ whereas that of the idler($\langle \hat{n}_c\rangle$)  increases from $0$ to $1$. The corresponding coefficients are given in Appendix B.

The positive contribution of thermal noise is explained as follows.
After interaction of the coherent state with thermal noise by a beam splitter and then tracing out the one of the output modes, there is no correlation in the final state, while the final mean photon number is a sum of the transmitted thermal noise and the reflected coherent state. However, for $N$-photon entangled states, there is correlation in the final state, while the final mean photon number is a product of the transmitted thermal noise and the coefficients of the reflected initial state.
Due to the final state correlation and quadratic terms of the final mean photon number, the transmitted thermal noise contributes to enhance the sensitivity and the SNR in the regime of $N_b\gtrsim 0.5$.
Therefore, $N$-photon entangled states can enhance not only the target sensitivity but also the target detection capability with increasing thermal noise.
Note that there is a similar behavior of enhancing phase sensitivity with increasing thermal noise\cite{Bagan,Oh19}.

\section{Summary and Discussion}
We made a connection between quantum-enhanced sensing and quantum illumination by starting with a scenario in which the phase sensitivity can be equivalent to the target sensitivity at low reflectivity. 
Including thermal noise and loss, we showed that the target sensitivity can be more enhanced with input of N-photon entangled states than with a TMSV state. 
Asymmetric entangled states exhibited better performance than the symmetric entangled states.
By using a photon-number difference measurement after a $50:50$ beam splitter, 
we found that the target sensitivity can be proportional to the target detection, 
resulting in both being enhanced with increasing thermal noise.

For the $N$-photon entangled states, there was a discrepancy between the QFI and the sensitivity with the error propagation relation($\Delta\eta$).
$\Delta\eta$ is enhanced with increasing thermal noise but the QFI decreases with it.
Since only the first-order field operation of the beam splitter was counted in the derivation\cite{Sanz} as $(\partial_{\eta}\rho_{\eta})|_{\eta=0}\approx tr_a[(\hat{a}^{\dag}\hat{b}-\hat{b}^{\dag}\hat{a}), |\psi\rangle_{ac}\langle \psi|\otimes\rho^{(b)}_{th}]$ in the QFI formula, 
we could not observe the effect of the high-order interference in the output state. However the $\Delta\eta$ included the higher-order field operations of the beam splitter, such that we could observe the effect of the high-order interference in the output state. 

Our SNR results show that thermal noise in a target detection can be beneficial when $\eta \ll 1$. 
It is therefore interesting to apply the SNR scenario to quantum ghost imaging\cite{SB12} and quantum-limited loss sensing\cite{Nair} which consider input entangled states in heavy noise environments.
As a further study, our measurement scheme can be modified even to utilize TMSV states. 

\begin{acknowledgments}
This work was supported by a grant to the Quantum Standoff Sensing Defense-Specialized Project funded by the Defense Acquisition Program Administration and the Agency for Defense Development.

\end{acknowledgments}

\section*{appendix A: Generation scheme of $N$-photon entangled states}
We propose a generation scheme of $N$-photon entangled states. Previously, two-photon entangled states have been generated with two beam splitters and two single-photon states\cite{two}. Both single-photon states are impinged on a beam splitter, and then one of the output modes goes through a beam splitter with no detection in an additional mode. 
Based on this idea, first, we describe how to produce four-photon entangled states in Fig. 4 (a). Two two-photon states are impinged on a beam splitter, and then each output mode goes through a beam splitter with no detection in an additional mode. After that, one of the output modes goes through another beam splitter with no detection in the additional mode. The final output state is given by
$|\text{4-photon}\rangle_{ac}=\sum^4_{n=0}C_{4-n,n}|4-n, n\rangle_{ac}$, and the success probability is derived as $P_s=\sum^4_{n=0}|C_{4-n,n}|^2$.
$C_{4-n,n}$ is a function of four different reflectivities of the beam splitters.
Note that the coefficients are not normalized.
For $N$-photon entangled states, in Fig. 4 (b), we inject $N-n$ and $n$ photons into a beam splitter, and then the output two-mode state goes through sequentially $N-1$ different beam splitters with no detection in the additional modes. The final output state is given by $|\text{N-photon}\rangle_{ac}=\sum^N_{n=0}C_{N-n,n}|N-n, n\rangle_{ac}$, and the success probability is derived as $P_s=\sum^N_{n=0}|C_{N-n,n}|^2$.
The coefficients are determined by the different reflectivities of the $N$ beam splitters. The success probability depends on the coefficients of the $N$-photon entangled state that we want to generate.
\begin{figure}[t]
\begin{center}
\vspace{0.05in}
\includegraphics[width=0.47\textwidth]{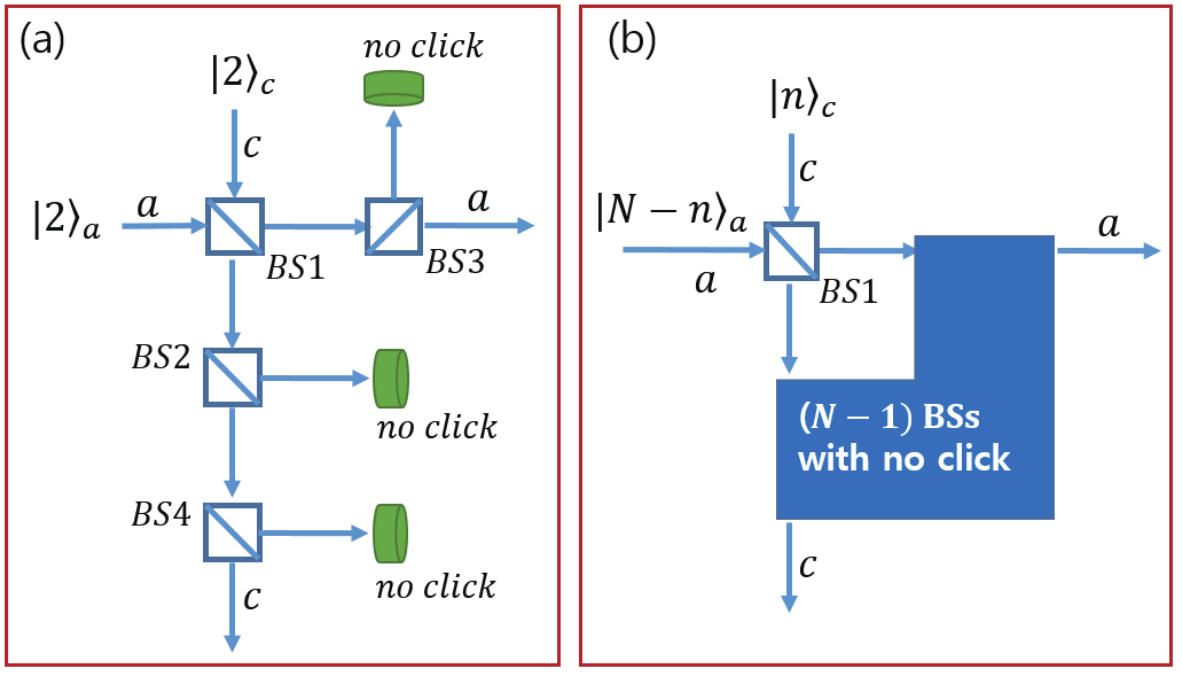}
\vspace{-0.1in}
\caption[0.6\textwidth]{Generation scheme: (a) four-photon entangled state, (b) $N$-photon entangled state.
BS stands for beam splitter. We do not inject any photon in the additional input modes.
There is no click event in the additional output modes.}
\label{Figure4}
\end{center}
\end{figure}

\section*{appendix B: Absolute values of coefficients optimized for QFI and SNR}
\begin{figure}[t]
\begin{center}
\centerline{\scalebox{0.31}{\includegraphics[angle=-90]{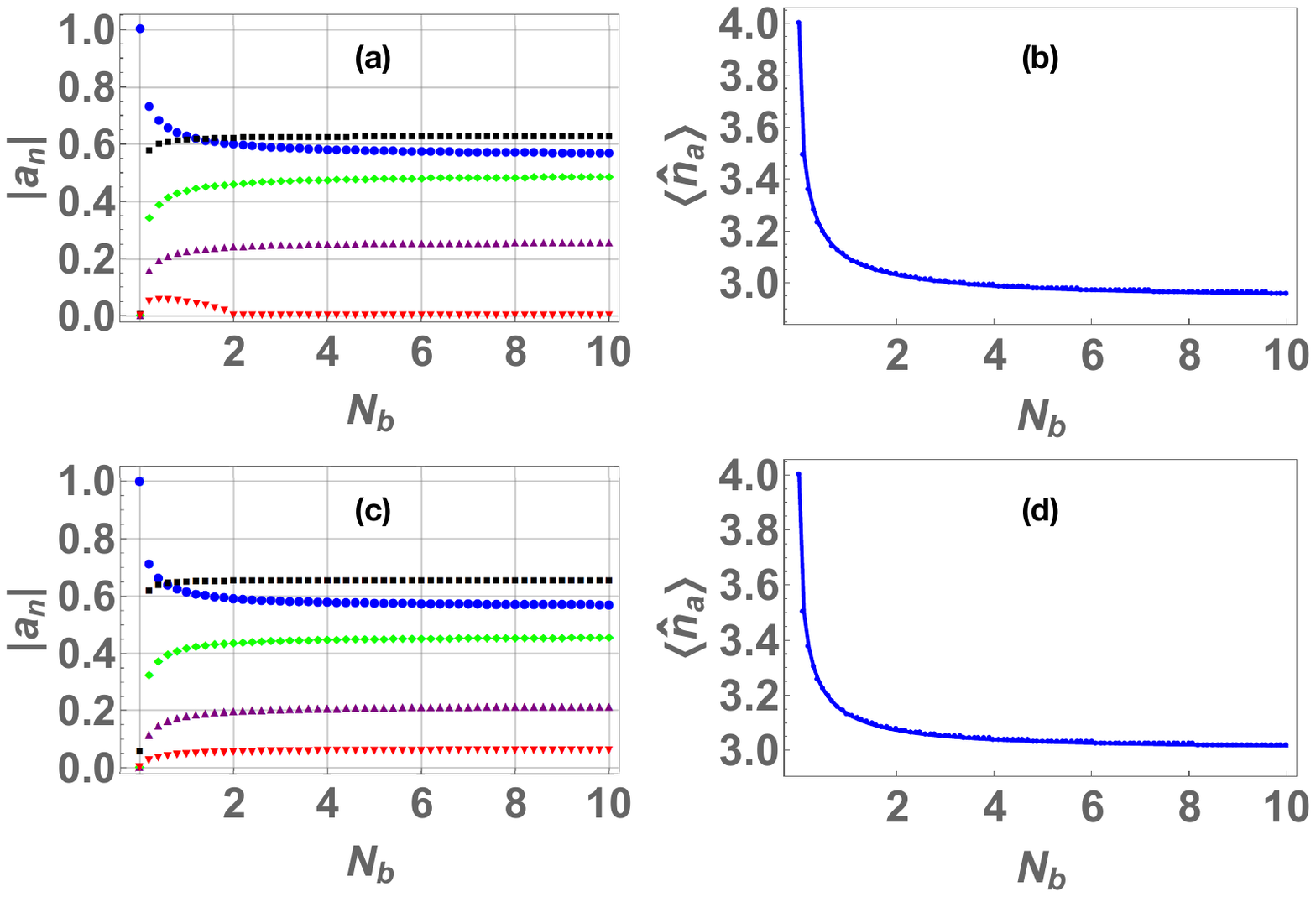}}}
\vspace{-0.1in}
\caption[0.6\textwidth]{Absolute values of coefficients optimized for QFI and SNR using a four-photon entangled state, as a function of the mean photon number of thermal noise($N_b$): $|a_0|$ (blue circles), $|a_1|$ (black squares), $|a_2|$ (green diamonds), $|a_3|$ (purple triangles), $|a_4|$ (red inverted triangles). 
(a) The rate of the coefficients for QFI, and (b) the corresponding mean photon number in the input mode $a$.
(c) The rate of the coefficients for SNR, and (d) the corresponding mean photon number in the input mode $a$. }
\label{Figure5}
\end{center}
\end{figure}

In Fig. 5, we show the absolute values of coefficients which are optimized for QFI and SNR using four-photon entangled states 
$\sum^4_{n=0}a_n|4-n,n\rangle_{ac}$. Note that $\Delta \eta$ has shown the same coefficients as the SNR.
At $N_b=0$, there exists only the coefficient $a_0$.  At $N_b>0$, the coefficients start with a descending order as $|a_0|>|a_1|>|a_2|>|a_3|>|a_4|$.  
With increasing $N_b$, the $|a_0|$ decreases while the absolute values of the other coefficients increase.
At $0<N_b<2$, the values of $|a_0|$ and $|a_1|$ cross over. In Fig. 5 (a), we note that the coefficient $|a_4|$ has non zero values with increasing $N_b$.
Based on the optimized coefficients, we calculate the mean photon number in the input mode $a$ as a function of $N_b$. 
With increasing $N_b$, the mean photon number $\langle\hat{n}_a\rangle$ decreases from $4$ to $3$, as shown in Fig. 5(b) and 5(d).  
It is almost saturated at $N_b=10$.
From the above optimization, we find that more photons are sent to the input mode $a$ than the other mode $c$ in Figs. 2 and 3.
This implies that, given the same input state energy, asymmetric entangled states can show better performance in the target sensitivity and the target detection.


\begin{thebibliography}{99}

\bibitem{giovannetti2004} V. Giovannetti, S. Lloyd, and L. Maccone, Nat. Photonics \textbf{5}, 222 (2011).

\bibitem{dowling2008} J.P. Dowling, Contemp. Phys. \textbf{49}, 125 (2008).

\bibitem{Stefano18} S. Pirandola, B.R. Bardhan, T. Gehring, C. Weedbrook, and S. Lloyd, Nat. Photonics \textbf{12}, 724 (2018).

\bibitem{BC94} S.L. Braunstein and C.M. Caves, \prl \textbf{72}, 3439 (1994).

\bibitem{Escher} B.M. Escher, R.D. de Matos Filho, and L. Davidovich, Nat. Physics \textbf{7}, 406 (2011).

\bibitem{Lloyd} S. Lloyd, Science \textbf{321}, 1463 (2008).

\bibitem{Tan} S.H. Tan, B.I. Erkmen, V. Giovannetti, S. Guha, S. Lloyd, L. Maccone, S. Pirandola, and J.H. Shapiro, \prl \textbf{101}, 253601 (2008). 

\bibitem{Shapiro} J.H. Shapiro, IEEE Aerosp. Electron. Syst. Mag. \textbf{35}, 8 (2020).

\bibitem{Genovese} E.D. Lopaeva, I. Ruo Berchera, I.P. Degiovanni, S. Olivares, G. Brida, and M. Genovese, \prl \textbf{110}, 153603 (2013).

\bibitem{Guha}S. Guha and B.I. Erkmen, \pra \textbf{80}, 052310 (2009).

\bibitem{Zheshen} Z. Zhang, S. Mouradian, F.N.C. Wong, and J.H. Shapiro, \prl \textbf{114}, 110506 (2015).

\bibitem{Quntao} Q. Zhuang, Z. Zhang, and J.H. Shapiro, \prl \textbf{118}, 040801 (2017).


\bibitem{SL09} J.H. Shapiro and S. Lloyd, New J. Phys. \textbf{11}, 063045 (2009).

\bibitem{Devi}A.R. Usha Devi and A.K. Rajagopal, \pra \textbf{79}, 062320 (2009).

\bibitem{Ragy} S. Ragy, I. Ruo Berchera, I. P. Degiovanni, S. Olivares, M. G. A. Paris, G. Adesso, and M. Genovese, J. Opt. Soc. Am. B \textbf{31}, 2045 (2014).

\bibitem{Zhang14} S.L. Zhang, J.S. Guo, W.S. Bao, J.H. Shi, C.H. Jin, X.B. Zou, and G.C. Guo, \pra \textbf{89}, 062309 (2014).

\bibitem{Shabir} S. Barzanjeh, S. Guha, C. Weedbrook, D. Vitali, J.H. Shapiro, and S. Pirandola, \prl \textbf{114}, 080503 (2015).

\bibitem{Sanz} M. Sanz, U. Las Heras, J.J. Garc\'ia-Ripoll, E. Solano, and R. Di Candia, \prl \textbf{118}, 070803 (2017).

\bibitem{Liu17} K. Liu, Q.-W. Zhang, Y.-J. Gu, and Q.-L. Li, \pra \textbf{95}, 042317 (2017).

\bibitem{Weedbrook}C. Weedbrook, S. Pirandola, J. Thompson, V. Vedral, and M. Gu, New. J. Phys.\textbf{18}, 043027 (2016).

\bibitem{Bradshaw}M. Bradshaw, S.M. Assad, J.Y. Haw, S.-H. Tan, P.K. Lam, and M. Gu, \pra \textbf{95}, 022333 (2017).

\bibitem{Zubairy}L. Fan and M.S. Zubairy, \pra \textbf{98}, 012319 (2018).

\bibitem{Stefano19} S. Pirandola, R. Laurenza, C. Lupo, and J.L. Pereira, npj Quantum Inf. \textbf{5}, 50 (2019).

\bibitem{Palma} G.De Palma and J. Borregaard, \pra \textbf{98}, 012101 (2018).

\bibitem{Yung}M.-H. Yung, F. Meng, X.-M. Zhang, and M.-J. Zhao, npj Quantum Inf. 6, 75 (2020).

\bibitem{Ray} S. Ray, J. Schneeloch, C.C. Tison, and P.M. Alsing, \pra \textbf{100}, 012327 (2019).

\bibitem{Sun}W.-Z. Zhang, Y.-H. Ma, J.-F. Chen, and C.-P. Sun, New J. Phys. \textbf{22}, 013011 (2020).

\bibitem{Ranjith} R. Nair and M. Gu, Optica \textbf{7}, 771(2020).

\bibitem{Shabir19} S. Barzanjeh,  S. Pirandola, D. Vitali, and J.M. Fink, Sci. Adv. \textbf{6}, eabb0451 (2020).

\bibitem{Sandbo}C.W. Sandbo Chang, A. M. Vadiraj, J. Bourassa, B. Balaji, and C.M. Wilson, Appl. Phys. Lett. \textbf{114}, 112601 (2019).

\bibitem{England}D.G. England, B. Balaji, and B.J. Sussman, \pra \textbf{99}, 023828 (2019).

\bibitem{Aguilar} G.H. Aguilar, M.A. de Souza, R.M. Gomes, J. Thompson, M. Gu, L. C. C\'eleri, and S. P. Walborn, \pra \textbf{99}, 053813 (2019).

\bibitem{Sussman} Y. Zhang, D. England, A. Nomerotski, P. Svihra, S. Ferrante, P. Hockett, and B. Sussman, \pra \textbf{101}, 053808 (2020).


\bibitem{Lee} S.-Y. Lee, Y.S. Ihn, and Z. Kim, \pra \textbf{101}, 012332 (2020).

\bibitem{BS} R.A. Campos, B.E.A. Saleh, and M.C. Teich, \pra  \textbf{40}, 1371 (1989).

\bibitem{Kim}M.S. Kim, J. Phys. B: At. Mol. Opt. Phys. \textbf{41}, 133001  (2008).

\bibitem{Lang13} M.D. Lang and C.M. Caves, \prl \textbf{111}, 173601 (2013).

\bibitem{Paris} M.G.A. Paris, Int. J. Quantum Inf. \textbf{7}, 125 (2009).

\bibitem{Rafal} R. Demkowicz-Dobrza\'nski, U. Dorner, B.J. Smith, J.S. Lundeen, W. Wasilewski, K. Banaszek, and I. A. Walmsley, \pra \textbf{80}, 013825 (2009).



\bibitem{Adesso} G. Adesso, D. Girolami, and A. Serafini, \prl \textbf{109}, 190502 (2012).


\bibitem{Adesso04} G. Adesso, A. Serafini, and F. Illuminati, \pra \textbf{70}, 022318 (2004).

\bibitem{Bagan} M. Aspachs, J. Calsamiglia, R. Mu\~noz-Tapia, and E. Bagan, \pra \textbf{79}, 033834 (2009).

\bibitem{Oh19} C. Oh, C. Lee, C. Rockstuhl, H. Jeong, J. Kim, H. Nha, and S.-Y. Lee, npj Quantum Inf. \textbf{5}, 10 (2019).

\bibitem{SB12} J.H. Shapiro and R.W. Boyd, Quantum Inf. Process. \textbf{11}, 949 (2012).

\bibitem{Nair} R. Nair, \prl \textbf{121}, 230801 (2018).

\bibitem{two} M. Kacprowicz, R. Demkowicz-Dobrza\'nski, W. Wasilewski, K. Banaszek, and I.A. Walmsley, Nat. Photonics \textbf{4}, 357 (2010).











\end{thebibliography}
\end{document}